# Monte Carlo Predictions of Proton SEE Cross-Sections from Heavy Ion Test Data*

Kai Xi (习凯)[1,2], Chao Geng (耿超)[3], Zhan-Gang Zhang (张战刚)[4], Ming-Dong Hou (侯明东)[1], You-Mei Sun (孙友梅)[1], Jie Luo (罗捷)[1], Tian-Qi Liu (刘天奇)[1], Bin Wang (王斌)[1,2], Bing Ye (叶兵)[1,2], Ya-Nan Yin (殷亚楠)[1,2], Jie Liu (刘杰)[1;1)]

[1] Institute of Modern Physics, Chinese Academy of Sciences, Lanzhou 730000, China

[2] University of Chinese Academy of Sciences, Beijing 100049, China

[3] Academy of Shenzhen State Microelectronics Co., LTD. Shenzhen 518057, China

[4] Science and Technology on Reliability Physics and Application of Electronic Component Laboratory, China Electronic Product Reliability and Environmental Testing Research Institute, Guangzhou 510610, China

**Abstract:** The limits of previous methods promote us to design a new approach (named PRESTAGE) to predict proton single event effect (SEE) cross-sections using heavy-ion test data. To more realistically simulate the SEE mechanisms, we adopt Geant4 and the location-dependent strategy to describe the physics processes and the sensitivity of the device. Cross-sections predicted by PRESTAGE for over twenty devices are compared with the measured data. Evidences show that PRESTAGE can calculate not only single event upsets induced by proton indirect ionization, but also direct ionization effects and single event latch-ups. Most of the PRESTAGE calculated results agree with the experimental data within a factor of 2-3.

## 1 Introduction

Energetic protons and heavy-ions (HIs) in space can induce single event effects (SEEs) such as single event upsets (SEU) and single event latch-ups (SEL) in electronic devices [1-5]. Those effects can significantly damage the on-orbit safety of satellites and spacecrafts. In the evaluation of SEE risk in space, error rates of the devices induced by both protons and HIs should be predicted. Protons are of more concern to researchers in certain applications due to their large flux in space. Although the most reliable way to calculate proton induced error rate for a device is to utilize experimentally measured proton SEE cross-sections, lots of attempts have been made to derive them from HI test data, denoted as SEE cross-sections varying with the LET (Linear Energy Transfer) of the testing ions. This is mainly due to the reason that HI tests are normally deemed to be essential and proton tests can lead to additional expenses for the researchers.

During the past two decades, several methods have been reported for deriving proton SEE sensitivity of devices from HI test data, such as the BGR [6], Petersen [7], PROFIT [8], Barak [9] and PROPSET [10] models. These models are able to make accurate predictions in some cases. However, they are not suitable for calculating cross-sections induced by proton direct ionization. Moreover, the errors of using them for SEL cross-section predictions could be higher than 2 orders [11]. The reason is that most of these models concerned only proton indirect ionization mechanism and employed numerous assumptions to simplify the analyses. In this work, a new method named PRESTAGE (PRoton-induced Effects Simulation Tool bAsed on GEant4) is proposed by adopting Geant4 [12-13] and the strategy of location-dependent sensitivity. Its calculation results for SEU and SEL effects of more than 20 devices are compared to the measured data.

*Supported by the National Natural Science Foundation of China (Grant Nos. 11179003, 10975164, 10805062 and 11005134).
1) E-mail: j.liu@impcas.ac.cn

## 2  Methodology

PRESTAGE uses Geant4 (9.6 version) as the basic platform to perform proton SEE simulations. Geant4 is an advanced Monte Carlo (MC) toolkit for simulations of energy deposition by particles passing through matter. The composition and utilization of PRESTAGE involve three procedures: device modeling, effect simulation, and cross-section calculation.

In the first procedure, the rectangular parallelepiped (RPP) method [14] is used for the device modeling. RPP model describes the cell of the device as a box with inside it a RPP-shaped sensitive volume (SV). Parameters indicating the SEE sensitivity of the device include the critical charge $Q_c$ and geometrical parameters of the SV (lateral dimensions $D_x$, $D_y$ and the thickness $T_{SV}$). These sensitive parameters are mainly derived from Weibull fitting [15] of the HI test data:

$$F(L) = \begin{cases} A\{1-\exp\{-[(L-L0)/W]^s\}\} & L > L0 \\ 0 & L < L0 \end{cases} \quad (1)$$

where $A$ is the limiting cross section, $L_0$ is the threshold LET in unit of MeV•cm$^2$/mg, $W$ is the width of the distribution, and $S$ is the shape parameter.

$D_x$ and $D_y$ are determined by: $D_x = D_y = \sqrt{A}$ [14]. $T_{SV}$ is closely related to the depth of the depletion area in junctions, and cannot be calculated from Weibull parameters. This value is normally obtained by experimental methods such as destructive physical analysis [15], pulse laser tests [16] or HI experiments [17]. $Q_c$ is the minimum amount of charge that must be collected by the device for an SEE to occur. Assuming all the charge that generated in and only in the SV is collected, we calculate $Q_c$ in unit of fC by:

$$Q_c = 10.36 \cdot L_c \cdot T_{SV} \quad (2)$$

where $T_{SV}$ is the thickness of the SV in unit of μm and $L_c$ is the critical LET. The constant 10.36 is calculated based on the facts that the energy needed to create an electron-hole pair in silicon is approximately 3.6 eV, and that the charge carried by an electron is about $1.602 \times 10^{-19}$ C. For a particular value of $T_{SV}$, $L_c$ directly determines the critical charge $Q_c$. According to Petersen et al.[10,18,19], the sensitivity to SEE is "location-dependent" within the SV and a reasonable way of deciding $L_c$ is by taking inverse of equation (1):

$$L_c = L_0 + W[-\ln(1-\frac{A_i}{A})]^{1/s} \quad (3)$$

where $L_0$, $W$, $A$, and $S$ are the Weibull parameters fitted from the HI test data, and $A_i$ is the top surface area of the sub-volume $i$ in the SV.

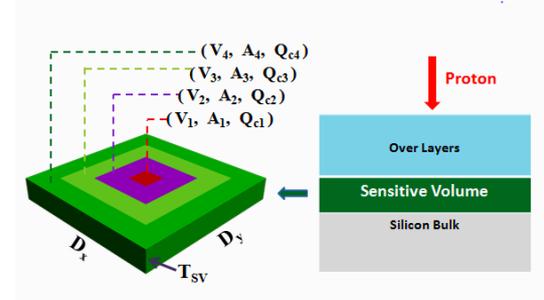

Fig. 1 Diagram of the device modeling. Only four sub-volumes in the SV are shown. Note that $V_N$ contains $V_{N-1}$, i.e. $V_2$ is the purple volume plus the red one.

Before the second procedure - effect simulation, the device modeling discussed above has to be realized in Geant4 (see Fig. 1). First, a box containing a RPP SV is defined representing the device under consideration. The SV is made of silicon with lateral dimensions $D_x$ and $D_y$. $SiO_2$ and other materials, representing over layers and surrounding of the device, are also added in the box. Second, the SV is divided into a number (more than 20) of sub-volumes, $V_1$, $V_2$ … $V_i$ … $V_N$, following Petersen's location-dependent sensitivity strategy. All these sub-volumes are concentric with each other and are placed in such a way that $V_1$ lies inside $V_2$, $V_2$ inside $V_3$…... and finally $V_{N-1}$ lies inside $V_N$. All these sub-volumes have the same thickness ($T_{SV}$) but increasing top surface areas denoted as $A_i$ ($1 \leq i \leq N$), with $A_N$ equals to the Weibull parameter A. $Q_{ci}$, the critical charge of $V_i$ with the top surface area $A_i$, is calculated from equations (2) and (3). All the physics processes required to run the simulation, including ion physics, hadron physics and lepton physics, etc., are contained in the Physics List [20]. Geant4 then automatically handles the ionization and nuclear reaction processes based on nuclear data libraries including the Evaluated Nuclear Data File (ENDF) in the USA.

Fig. 2 shows the simplified flow chart of the effect simulation process. A number of protons ($N_t$) with the same energy hit normally and randomly at the upper surface of the box. For one proton penetration, if the charge deposited in $V_i$ exceeds $Q_{ci}$, the sub-volume $V_i$ is marked. An SEE is deemed to be triggered if one or more sub-volumes are marked after the proton penetration process.

The last process, cross-section calculation, occurs when the effect simulation of $N_t$ protons is finished. The total number of triggered SEE, $N_e$ is counted and the SEE cross-section can be calculated by the relation:

$$\sigma_p = \frac{N_e}{N_t} \cdot A_b \quad (4)$$

Where $\sigma_p$ is the calculated SEE cross-section induced by proton and $A_b$ is the upper surface area of the box. To evaluate the validity of PRESTAGE, calculations were performed with several devices that had been tested under both proton and HI beams.

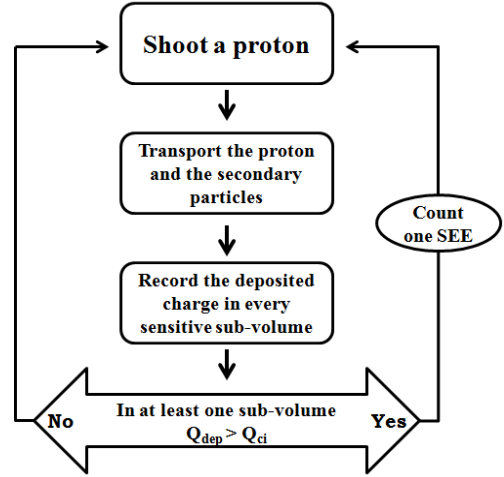

Fig. 2 Flow chart of the effect simulation process.

## 3 Calculation results for SEU effects

It is traditionally believed that protons, with low LET values, cause SEEs mainly by nuclear reaction (indirect ionization) mechanism. In recent years, however, various researches have reported that proton direct ionization can also induce SEU in certain nano-scale devices, and the corresponding SEU cross-section could be 3-4 orders of magnitude higher than the saturation cross-section induced by nuclear reaction mechanism [4-5]. In this section, PRESTAGE calculations for SEU effects induced by both direct and indirect ionizations are performed.

### 3.1 Proton indirect ionization

The validity of PRESTAGE for prediction of SEU effects induced by proton indirect ionization was studied by calculating cross-sections for the configuration memory of the Xilinx Virtex-II X-2V1000 [21]. This device is a Field Programmable Gate Array (FPGA) fabricated in 0.15 μm bulk CMOS technology. PRESTAGE input values (listed in Table 1), such as the Weibull parameters of the HI test data and the $T_{SV}$, had been reported previously in literatures [10, 21]. A 2-μm $SiO_2$ layer above the SV was assumed as a representative of the passivation layer.

Table 1. Device names and the published PRESTAGE input parameters.

| Device | $L_0$ /(MeV cm$^2$/mg) | A /(cm$^2$/bit) | W | S | $T_{SV}$ /μm | Ref. |
|---|---|---|---|---|---|---|
| Xilinx Virtex-II FPGA_Config Mem | 1.00 | 4.37E-08 | 33.0 | 0.80 | 0.15 | 10,21 |
| 90-nm SRAM | 0.33 | 5.76E-08 | 22.8 | 1.07 | 0.10 | 10,22 |

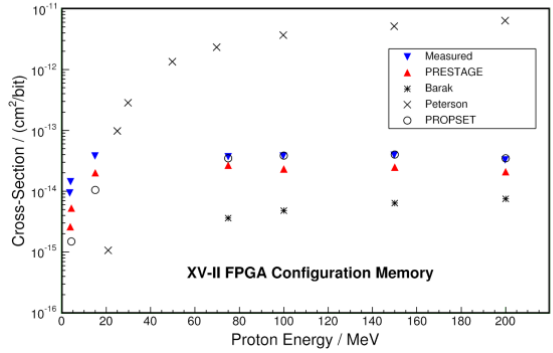

Fig. 3. Comparison of proton induced SEU cross-sections for the Xilinx Virtex-II FPGA configuration memory predicted by PRESTAGE and other models with measured results. Calculation results of Petersen's model are taken from Koga et al. [21].

Fig. 3 shows the result of PRESTAGE calculated upset cross-sections at various incident proton energies for the Virtex-II FPGA configuration memory in comparison with the measured data [21]. Predictions using Barak, Petersen and PROPSET models are also shown in the figure. Barak and Petersen models are semi-empirical methods. To simplify the analysis and get an analytical solution, SVs of devices in these models are usually assumed to be a dot or an infinite volume [6, 7, 9] while considering the influential energy deposited by the recoils. In reality, the situation is much more complex. PROPSET and PRESTAGE adopt the location-dependent sensitivity strategy [10, 18, 19] as a more sophisticated way to describe the SV and its susceptibility to radiations. As a result, the predictions of PRESTAGE and PROPSET are shown in better agreement with the measured data (see Fig. 3), and most of the predicted cross-sections agree with the experimental data within a factor of two.

PRESTAGE differs with PROPSET. When launching an event, PROPSET immediately generates nuclear reactions within the SV, whereas PRESTAGE realistically tracks the proton from the top surface of the device, through the over-layers, and into the SV. This feature enables PRESTAGE to simulate not only proton indirect ionization but also direct ionization effects.

### 3.2  Proton direct ionization

A SRAM bit cell fabricated in commercial 90-nm process was used to evaluate the validity of PRESTAGE in predictions of SEU effects induced by proton direct ionization. The corresponding HI tested data was taken from Cannon et al. [22]. In this simulation, the passivation layers of the device were modeled as a 4.9-μm-thick polyimide layer above an 8.9-μm-thick oxide layer [22]. Other PRESTAGE input variables, such as the $T_{SV}$ and the Weibull fitting parameters, are listed in Table 1. Fig. 4 shows the calculated results by PRESTAGE and other methods at different incident proton energies compared to the measured cross-sections reported in [22]. Unlike upsets of the Virtex-II FPGA that results entirely from nuclear reaction mechanism, upsets of the 90-nm SRAM can also be induced by proton direct ionization. As can be seen in Fig. 4, the cross-sections calculated by PRESTAGE agree with the experimental data in both direct and indirect ionization regions.

Predicted values using Barak and Petersen models are also displayed in Fig. 4. Referring to our data the measured results are more closely matching the Barak curve rather than the Petersen curve in the nuclear reaction range. Neglecting direct ionization mechanism, they both fail to accurately predict cross-sections at low proton energies.

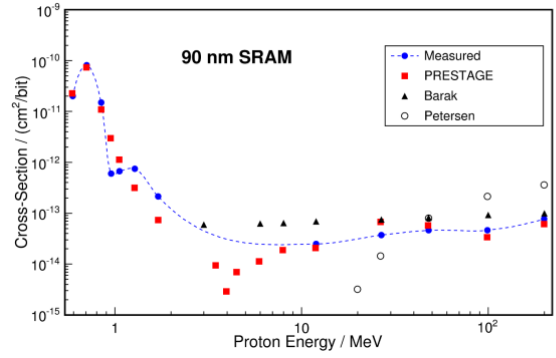

Fig. 4. Comparison of PRESTAGE predicted SEU cross-sections for the 90-nm SRAM [22] induced by both proton direct and indirect ionizations, with measured data. Calculation results from Barak and Petersen models are also shown in the figure.

### 3.3 Other results for SEU effect calculations

A number of parts that had been tested for SEU under heavy ion and proton beams are listed in Table 2. The fitted Weibull parameters and the nominal SV thicknesses had been reported [10]. PRESTAGE was used to calculate the saturation cross-section ($\sigma_{sat}$) induced by protons with incident energy of 200 MeV for these parts. The calculated and measured $\sigma_{sat}$ are compared in Table 3, and their ratio (calculated $\sigma_{sat}$ divided by measured $\sigma_{sat}$) is also given.

Table 2. Information of parts susceptible to SEU

| Part # | Part ID | L0 /(MeV cm²/mg) | A/ (cm²/bit) | W | S | $T_{SV}$ /μm |
|---|---|---|---|---|---|---|
| 1 | IBM_16MEG | 1.7 | 7.74E-09 | 20 | 3 | 0.2 |
| 2 | 01G9274 | 1.6 | 2.30E-08 | 28 | 3.25 | 0.2 |
| 3 | LUNA_C | 3.2 | 8.93E-09 | 14 | 3 | 0.2 |
| 4 | 1601_EPI | 2.75 | 6.25E-06 | 30 | 1.5 | 1.0 |
| 5 | OW62256 | 2.9 | 1.90E-06 | 14 | 2.3 | 1.0 |
| 6 | SMJ44100 | 1.39 | 4.76E-07 | 15 | 1.21 | 1.0 |
| 7 | 62256 | 1.6 | 2.44E-06 | 20 | 1.65 | 1.0 |
| 8 | MT4C1004C | 1.49 | 3.09E-07 | 20 | 1.2 | 1.0 |
| 9 | HM6116 | 4.2 | 4.12E-06 | 7.9 | 2.5 | 1.0 |
| 10 | 62832H | 3.4 | 3.80E-07 | 20 | 1.5 | 1.0 |
| 11 | TC514100Z-10 | 0.86 | 5.00E-07 | 18 | 1.15 | 1.0 |

| 12 | HM_65656 | 1.5 | 4.20E-07 | 12 | 1.75 | 1.0 |
| 13 | MB814100_10PSZ | 1.15 | 7.62E-07 | 15 | 1.35 | 1.0 |
| 14 | HYB514100J_10 | 0.86 | 5.00E-07 | 14 | 1.1 | 1.0 |
| 15 | 93L422-Fairchild | 0.6 | 2.60E-05 | 4.4 | 0.7 | 2.0 |

The part numbers in this paper, the part IDs, the published Weibull parameters and nominal SV thicknesses are listed.

Table 3. Measured and calculated results of parts that susceptible to SEU.

| Part # | Measured $\sigma_{sat}$ /(cm$^2$/Device) | PRESTAGE calculated $\sigma_{sat}$ /(cm$^2$/Device) | PROPSET calculated $\sigma_{sat}$ /(cm$^2$/Device) | PRESTAGE Ratio Calc/Measured | PROPSET Ratio Calc/Measured |
|---|---|---|---|---|---|
| 1 | 2.12E-08 | 9.20E-09 | 2.85E-08 | 0.43 | 1.81 |
| 2 | 4.19E-09 | 4.70E-09 | 1.68E-08 | 1.11 | 4.01 |
| 3 | 2.12E-08 | 1.70E-08 | 7.76E-08 | 0.80 | 3.36 |
| 4 | 9.00E-08 | 1.28E-07 | 1.75E-07 | 1.43 | 1.94 |
| 5 | 8.70E-08 | 2.69E-07 | 2.71E-07 | 3.09 | 3.12 |
| 6 | 7.00E-07 | 1.01E-06 | 2.08E-06 | 1.44 | 2.90 |
| 7 | 1.47E-07 | 3.90E-07 | 4.46E-07 | 2.60 | 3.15 |
| 8 | 2.94E-07 | 8.69E-07 | 9.41E-07 | 2.95 | 3.20 |
| 9 | 4.59E-08 | 6.24E-08 | 8.98E-08 | 1.36 | 1.96 |
| 10 | 2.89E-08 | 4.87E-08 | 4.06E-08 | 1.68 | 1.41 |
| 11 | 1.00E-06 | 1.60E-06 | 2.07E-06 | 1.60 | 2.07 |
| 12 | 2.98E-08 | 8.03E-08 | 1.17E-07 | 2.70 | 3.91 |
| 13 | 6.90E-07 | 2.86E-06 | 3.45E-06 | 4.14 | 5.00 |
| 14 | 1.46E-06 | 2.07E-06 | 2.72E-06 | 1.41 | 1.86 |
| 15 | 1.42E-07 | 9.80E-08 | 1.10E-07 | 0.69 | 0.78 |

The part numbers, the measured proton saturation cross-sections, the PRESTAGE and PROPSET [10] calculated saturation cross-sections, and their ratios to the measured data are listed.

Table 3 presents the measured and PRESTAGE calculated results for devices listed in Table 2. Predictions of PROPSET [10] are also provided as a reference. As can be seen, the PRESTAGE calculated $\sigma_{sat}$ agrees with the measured $\sigma_{sat}$ within a factor of three for most of the cases. Moreover, PRESTAGE tends to moderately over predict the limiting cross-section, which is favorable for a conservative estimation of risk in space.

## 4 Calculation results for SEL effects

SEL effects used to be difficult problems to solve, because the corresponding effective depths of SV are usually comparable to the ranges of the recoils generated from nuclear reactions between the proton and silicon [9]. Models such as Barak or BGR could be used to give analytical solutions for very small or large SVs, as they simplify the SV ether to a dot or to an infinite volume. But in cases of SEL, where the SV dimensions are comparable to the recoil ranges, calculation errors of these models could be unacceptable [9, 11]. In this section, SEL cross-sections predicted by PRESTAGE are presented.

Table 4. Information of parts susceptible to SEL

| Part ID | L0 | A/ | W | S | $T_{SV}$ /μm |
|---|---|---|---|---|---|

|  | /(MeV cm$^2$/mg) | (cm$^2$/bit) |  |  |  |
|---|---|---|---|---|---|
| K-5 | 0.10 | 2.90E-01 | 54.1 | 3.34 | 3.5 |
| 32C016 | 1.20 | 5.00E-02 | 42.3 | 2.51 | 13 |
| LSI-64811 | 1.4 | 2.00E-01 | 35.8 | 4.44 | 15 |
| NCE4464JPL | 1.00 | 1.20E-01 | 18.5 | 3.78 | 15 |
| HM65162-85 | 2.30 | 2.00E-02 | 8.48 | 3.33 | 18 |

The part IDs, the published Weibull parameters, and the published nominal SV thicknesses of the parts are listed.

Table 5. Measured and calculated results of parts that susceptible to SEL.

| Part ID | Measured $\sigma_{pr}$ /(cm$^2$/Device) | PRESTAGE calculated $\sigma_{pr}$ /(cm$^2$/Device) | Ratio Calc/Measured |
|---|---|---|---|
| K-5 | 6.60E-09 | 8.25E-09 | 1.25 |
| 32C016 | 1.00E-09 | 1.65E-09 | 1.65 |
| LSI-64811 | 1.70E-11 | 5.48E-11 | 3.22 |
| NCE4464JPL | 2.00E-10 | 3.60E-10 | 1.80 |
| HM65162-85 | 4.00E-10 | 8.40E-10 | 2.10 |

The part ID of the devices, the measured and PRESTAGE calculated saturation cross-sections and their ratios are listed.

Table 4 presents the information of five semiconductor devices which were susceptible to SEL. The Weibull parameters were reported by Normand [23], and the corresponding $T_{SV}$ values were revealed by Johnston et al. [24]. PRESTAGE was used to calculate the SEL cross-sections induced by 200 MeV protons based on the information provided in Table 4. Table 5 presents the comparison between the calculated results and the measured data [23]. The ratio in the last column is the PRESTAGE calculated cross-section divided by the measured one. Good agreement is observed in the comparisons.

## 5  Conclusion

A Monte Carlo method - PRESTAGE was proposed to calculate SEE cross-sections induced by protons. Distinguished from previous methods, PRESTAGE describes the physics processes and susceptibility of the device in a more realistic way by adopting Geant4 and the strategy of location-dependent sensitivity in SV. It is able to predict not only SEUs induced by proton indirect ionization, but also SELs and direct ionization effects. We used PRESTAGE to calculate SEE cross-sections of more than 20 devices triggered by protons with different energies. Most of the calculated results agreed with the measured data within a factor of 2-3. The agreement indicates that PRESTAGE provides a reliable way of predicting proton SEE sensitivity, especially in situations where proton beam tests are not available.

## References

1  Lei Z F, Guo H X, Zeng C et al, Chin. Phys. B, **24**: 056103 (2015)
2  Xiao Y, Guo H X, Zhang F Q et al, Chin. Phys. B, **23**: 118503（2014）
3  Chen S M and Chen J J，Chin. Phys. B，**21**：016104（2012）
4  Heidel D F, Marshall P W, LaBel K A et al, IEEE Trans. Nucl. Sci., **55**: 3394 (2008)
5  Schwank J R, Shaneyfelt M R, Cavrois V F et al, IEEE Trans. Nucl. Sci., **59**:1197 (2012)
6  Normand E, IEEE Trans. Nucl. Sci., **45**: 2904 (1998)
7  Petersen E L, IEEE Trans. Nucl. Sci., **39**: 1600 (1992)
8  Calvel P, Barillot C, Lamothe P et al, IEEE

Trans. Nucl. Sci., **43:** 2827 (1996)

9  Barak J, IEEE Trans. Nucl. Sci., **53**: 3336 (2006)

10 Foster C C, O'Neill P M and Kouba C K, IEEE Trans. Nucl. Sci., **53**: 3494 (2006)

11 Edmonds L D, IEEE Trans. Nucl. Sci., **47**: 1713 (2000)

12 Agostinelli S, Allisonas J, Amakoe K et al, Nuclear Instruments and Methods in Physics Research A **506**: 250 (2003)

13 Allison J, Amako K, Apostolakis J et al. IEEE Trans. Nucl. Sci., **53**: 270 (2006)

14 ECSS-E-ST-10-12C: Methods for the calculation of radiation received and its effects and a policy for design margins. 2008.

15 Schwank J R, Shaneyfelt M R and Dodd P E, IEEE Trans. Nucl. Sci., **60**: 2074 (2013)

16 Jones R, Chugg A M, Jones P et al, 6th European Conference on Radiation and Its Effects on Components and Systems, September 2001 p. 380

17 Inguimbert C, Duzellier S, Ecoffet R et al, IEEE Trans. Nucl. Sci. **47**: 551 (2000)

18 Petersen E L, Pickel J C, Adams J H et al, IEEE Trans. Nucl. Sci., **39**: 1577 (1992)

19 Petersen E L, Pouget V, Massengill L W et al, IEEE Trans. Nucl. Sci., **52**: 2158 (2005)

20 http://www.slac.stanford.edu/comp/physics/geant4/slac_physics_lists/G4_Physics_Lists.html

21 Koga R, George J, Swift G et al, IEEE Trans. Nucl. Sci., **51**: 2825 (2004)

22 Cannon E H, Holmen M C, Wert J et al, IEEE Trans. Nucl. Sci., **57**: 3493 (2010)

23 E. Normand, IEEE Trans. Nucl. Sci., **51**: 3494 (2004)

24 Johnston A H and Swift G M, IEEE Trans. Nucl. Sci., **44**: 2367 (1997)